\documentclass[conference]{IEEEtran}
\IEEEoverridecommandlockouts

\usepackage{cite}
\usepackage{amsmath,amssymb,amsfonts}
\usepackage{algorithmic}
\usepackage{graphicx}
\usepackage{textcomp}
\usepackage{xcolor}
\def\BibTeX{{\rm B\kern-.05em{\sc i\kern-.025em b}\kern-.08em
    T\kern-.1667em\lower.7ex\hbox{E}\kern-.125emX}}

\usepackage{booktabs}
\usepackage{balance}

\usepackage[hidelinks]{hyperref}
\hypersetup{pdfauthor={Zechun Zhou, Zhiyong Zheng, Xiaomin Zhang, Lin Ye, Han Li, Jingwei Sun, Guangzhong Sun, and Jing Li},pdftitle={AscMLA: Unlocking FP4 Efficiency for Long-Context MLA Decoding through Asymmetric Scaling},pdfsubject={FP4 efficiency for long-context MLA decoding},pdfkeywords={AscMLA,MLA,FP4,KV cache,asymmetric scaling,GPU kernels}}
\newcommand{\Qfour}{\operatorname{Q}_{4}}
\newcommand{\Qeight}{\operatorname{Q}_{8}}
\newcommand{\softmax}{\operatorname{softmax}}

\begin{document}
\title{AscMLA: Unlocking FP4 Efficiency\\for Long-Context MLA Decoding\\through Asymmetric Scaling}
\author{%
\IEEEauthorblockN{Zechun Zhou\textsuperscript{1*},
Zhiyong Zheng\textsuperscript{2},
Xiaomin Zhang\textsuperscript{2},
Lin Ye\textsuperscript{2},\\
Han Li\textsuperscript{1},
Jingwei Sun\textsuperscript{1\textdagger},
Guangzhong Sun\textsuperscript{1\textdagger}, and
Jing Li\textsuperscript{1}}
\IEEEauthorblockA{\textsuperscript{1}University of Science and Technology of China\quad
\textsuperscript{2}Kuaishou Technology\\
\{zhouzechun, hanli06\}@mail.ustc.edu.cn\quad \{sunjw, gzsun, lj\}@ustc.edu.cn\\
\{zhengzhiyong, zhangxiaomin03, yelin03\}@kuaishou.com}
\thanks{\hspace*{-\parindent}\makebox[.5em][l]{\textsuperscript{*}}\,Work done during an internship at Kuaishou Technology.\par
\noindent\makebox[.5em][l]{\textsuperscript{\textdagger}}\,Correspondence to: sunjw@ustc.edu.cn and gzsun@ustc.edu.cn.}
}
\maketitle

\begin{abstract}
Long-context decoding repeatedly reads a growing KV cache. FP4 reduces cache storage and traffic but needs block scaling to handle its limited dynamic range. Applying FP4 to multi-head latent attention (MLA) requires reconciling different K/V scaling requirements on a shared latent. Existing approaches address this mismatch through dequantization or separate encodings, adding execution or storage overhead. We present AscMLA, which combines asymmetric K/V scaling with pipelined scale delivery to process a single FP4 payload directly. AscMLA uses 38.2\% less cache memory than FP8 and achieves geometric-mean speedups of 1.35$\times$ over TRT-LLM's FP8 and 1.26$\times$ over CAKE's NVFP4 MLA kernels.
\end{abstract}

\begin{IEEEkeywords}
FP4 KV cache, MLA kernels, LLM serving
\end{IEEEkeywords}

\section{Introduction}
Long-context autoregressive decoding places growing demands on memory bandwidth and capacity. Each generation step rereads the KV history, while concurrent requests keep their histories resident. Reducing cache precision can lower storage and increase concurrency~\cite{kivi}. FP8 attention kernels already exploit this opportunity~\cite{flashinfer}. Moving to FP4 promises further storage savings and higher matrix throughput.

To limit quantization error from FP4's narrow dynamic range, formats such as NVFP4~\cite{nvfp4} and MXFP4~\cite{mxformats} use block scaling, with one scale per block of values. Native block-scaled matrix instructions apply these scales along the reduction dimension~\cite{mpfa}. This requirement poses a challenge for multi-head latent attention (MLA)~\cite{deepseekv2}. MLA reduces KV cache storage by sharing a compressed latent between K and V~\cite{deepseekv3}. Query--key multiplication (QK) reduces over latent channels, whereas probability--value multiplication (PV) reduces over tokens. Channel-block scales align with the reduction dimension in QK but lie along an output dimension in PV, so the native PV instruction cannot apply the same block scales directly.

FP8 MLA kernels, including FlashInfer's CuTe DSL implementation~\cite{flashinfer} and TRT-LLM's implementation~\cite{trtllm}, have provided production-ready decoding solutions in serving systems such as vLLM~\cite{vllm} and SGLang~\cite{sglang}. With FP4 MLA support still limited, a naive way to extend these FP8 implementations to FP4 caches is to dequantize latents to FP8 in a separate kernel before attention. CAKE's NVFP4 MLA kernel~\cite{cake} goes further by keeping QK in native FP4 and using QMUL4 instructions to efficiently dequantize V to FP8 within attention. Fusing V dequantization into attention reduces conversion overhead, but V conversion remains a bottleneck. SageAttention3~\cite{sageattention3} is an FP4 method for multi-head attention (MHA)~\cite{transformer} that separately encodes K and V for their respective matrix products. Adapting this design to MLA avoids V dequantization but duplicates the cached latent. The challenge is to use the FP4 values directly in PV without giving up the storage advantage of MLA’s shared latent.

We present AscMLA, a representation and kernel co-design for efficient MLA decoding with FP4 KV caches. Our key observation is that sharing FP4 codes does not require sharing scales. We choose one V scale shared across each token's channels so that the kernel can fold it into the corresponding softmax probability~\cite{snapmla}, leaving V in FP4. Because this scale cannot adjust channel blocks independently, AscMLA quantizes the latent for V first to obtain the shared codes. With those codes fixed, it fits one K scale per 16-channel block to enable native NVFP4 matrix multiply-accumulate (MMA) for QK. For PV, the kernel represents the scaled softmax probabilities in FP8 and multiplies them by the shared FP4 codes using MXFP8--MXFP4 MMA.

Efficient execution further requires delivering these scales without stalling the asynchronous pipeline. K scales must remain valid until MMA completes, whereas V scales are consumed during probability conversion. AscMLA matches buffering to these distinct lifetimes. K-scale multibuffering removes wait-before-overwrite dependencies between matrix operations, while V-scale multistage prefetching overlaps token-scale delivery with score production.

We evaluate AscMLA against five representative baselines in terms of storage, performance, and accuracy. Across all forty benchmark configurations, AscMLA achieves geometric-mean speedups of 1.35$\times$ over TRT-LLM's FP8 and 1.26$\times$ over CAKE's NVFP4 MLA kernels, while also using 38.2\% less KV cache memory than FP8. Attention-output tests and GPQA-Diamond~\cite{gpqa} evaluations of Kimi-K2.6~\cite{kimik26} and Kimi-K3~\cite{kimik3} quantify the accuracy tradeoff. In end-to-end Kimi-K2.6 serving, AscMLA delivers 18.8\% higher throughput than the TRT-LLM FP8 baseline at the highest evaluated load while meeting both latency targets.

The main contributions of our work are as follows:
\begin{itemize}
 \item We propose a shared-payload FP4 representation for MLA latents with asymmetric K/V scaling.
 \item We design hardware-efficient FP4 MLA decoding with native low-bit MMA and pipelined scale delivery.
 \item We evaluate AscMLA against five representative baselines, assessing accuracy and demonstrating improved kernel efficiency and serving throughput.
\end{itemize}

\section{Background and Motivation}
\subsection{Challenges in FP4 MLA Decoding}
MLA reduces KV storage by sharing a compressed latent between K and V~\cite{deepseekv2}. For $T$ cached tokens, let $X\in\mathbb{R}^{T\times d}$ denote this latent and $R\in\mathbb{R}^{T\times r}$ the separate K-side rotary position embedding (RoPE) component~\cite{roformer}. Absorbing the K/V projections into the query and output transforms gives
\begin{equation}
 \begin{aligned}
 L&=\beta(Q_cX^\top+Q_rR^\top),\\
 P&=\softmax(L),\qquad Y=PX.
 \end{aligned}
 \label{eq:mla}
\end{equation}
Here $Q_c$ and $Q_r$ are the content and RoPE queries for $H$ local heads, and $\beta$ is the attention factor. Both products share $X$, which has no KV-head axis.

\begin{figure}[t]
\centering
\includegraphics[width=\linewidth]{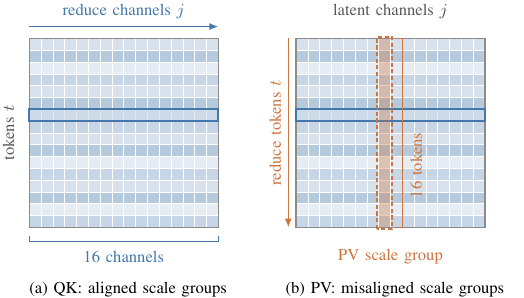}
\caption{NVFP4 scale-group mismatch in MLA. (a) The 16 channels in the blue row share one cached scale, which native QK MMA can use directly. (b) The 16 tokens in the orange column have separate cached scales, whereas native PV MMA requires one shared scale for the entire group.}
\label{fig:orientation}
\end{figure}

Block scaling complicates this sharing because the products reduce along different axes. NVFP4 represents the latent as $\widehat X_{t,j}=g\,s_{t,c(j)}Z_{t,j}$, with E2M1 codes $Z$, E4M3 scales $s$, a global factor $g$, and 16-channel groups indexed by $c(j)=\lfloor j/16\rfloor$~\cite{nvfp4}. These groups align with QK's channel reduction, allowing native block-scaled MMA to apply the scales along GEMM-K~\cite{mpfa}. In PV, however, the reduction is over tokens:
\begin{equation}
 (P\widehat X)_{h,j}
 =g\sum_t P_{h,t}\,s_{t,c(j)}Z_{t,j}.
 \label{eq:pvblock}
\end{equation}
Fig.~\ref{fig:orientation} shows that a QK group shares one cached scale across 16 channels, whereas a PV group spans 16 tokens with independent cached scales. Because $s_{t,c(j)}$ depends on both token and output channel group, it can neither be absorbed into a probability shared across channels nor factored outside the token sum. Transposing the codes alone cannot remove this mismatch~\cite{bitdecodingfp4}, so the QK encoding cannot be reused unchanged for native block-scaled PV.

\subsection{Memory Traffic Analysis}
\label{sec:execution-cost}
CAKE's NVFP4 MLA kernel~\cite{cake} combines native FP4 QK with efficient QMUL4-based V dequantization. Its PV MMA still reads an expanded FP8 operand from shared memory (SMEM), while QK block scales and MMA accumulators reside in tensor memory (TMEM)~\cite{fa4}. The converted V must therefore be written to SMEM before PV, even when dequantization is fused into attention.

We profile TRT-LLM's FP8 kernel, CAKE's NVFP4 kernel, and AscMLA with Nsight Compute under matched workloads. At $B=32$, $T=64$K, and $H=128$, their DRAM reads are 1.211, 0.740, and 0.748 GB, respectively. CAKE's NVFP4 kernel and AscMLA reduce DRAM reads by similar amounts relative to the FP8 method, but only CAKE writes dequantized V to SMEM before PV. CAKE's SMEM store instructions request 1.384 GB, compared with 0.285 GB each for TRT-LLM's FP8 kernel and AscMLA. The additional 1.099 GB is comparable to the 1.074 GB expanded V operand. CAKE's NVFP4 kernel thus reads 1.1\% less data from DRAM than AscMLA but requests 4.85$\times$ as many SMEM store bytes.

Efficient fused dequantization therefore preserves FP4's DRAM savings but still incurs the SMEM cost of V expansion. The additional SMEM traffic motivates a shared FP4 representation that both products can consume directly, with scale delivery designed to maintain asynchronous execution.

\section{Design and Implementation}
\label{sec:design}
\subsection{Overview}
\begin{figure*}[t]
\centering
\includegraphics[width=\linewidth]{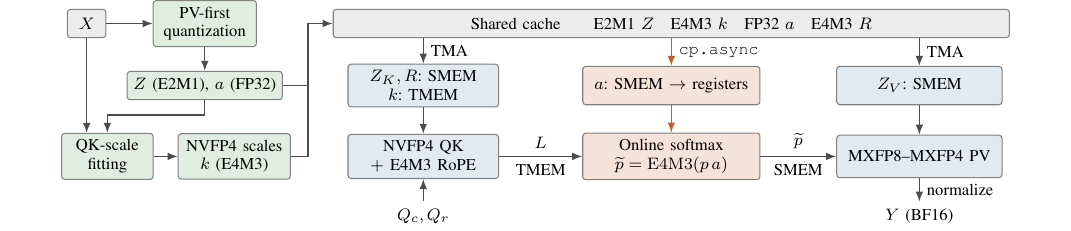}
\caption{Overview of AscMLA. PV-first quantization produces a shared FP4 payload $Z$ and per-token V scales $a$. QK-scale fitting derives block K scales $k$ without changing $Z$. The payload is loaded into K/V SMEM layouts $Z_K$ and $Z_V$ for native NVFP4 QK and MXFP8--MXFP4 PV. V scales are folded into probabilities before PV.}
\label{fig:design}
\end{figure*}

Fig.~\ref{fig:design} summarizes AscMLA's design. The cache writer constructs FP4 codes with per-token V scales, then fits channel-block K scales to those codes. QK uses native NVFP4 MMA, while PV folds V scales into probabilities and consumes the same FP4 payload directly. The decode pipeline keeps K scales available until the corresponding QK MMA completes and prefetches V scales before the probabilities are ready.

\subsection{Asymmetrically Scaled FP4 KV Cache}
\label{sec:representation}
\label{sec:construction}
In AscMLA, the cache pairs one E2M1 latent payload $Z$ with an E4M3 K scale $k$ per 16 channels and an FP32 V scale $a$ per token. These scale views define two approximations of the same latent:
\begin{equation}
 \widehat K_{t,j}=k_{t,c(j)}Z_{t,j},\qquad
 \widehat V_{t,j}=a_tZ_{t,j}.
 \label{eq:shared}
\end{equation}
The K view follows NVFP4 block16 semantics. Scales are nonnegative and global factors are one. The K-side RoPE component $R$ remains separate in E4M3.

Unlike the block scales in Eq.~\eqref{eq:pvblock}, $a_t$ does not depend on the output channel. It can therefore be folded into the probability, as in per-token-scaled FP8 MLA~\cite{snapmla}:
\begin{equation}
 P\widehat V=P\operatorname{diag}(a)Z
             =\bigl(P\operatorname{diag}(a)\bigr)Z.
 \label{eq:valuefold}
\end{equation}
This identity replaces scaling $d$ V elements per token with scaling one probability per head and token.

K scales can adjust channel blocks independently, whereas V has only one scale per row. K-oriented codes can therefore lose relative block amplitudes that a single V scale cannot recover. AscMLA uses PV-first construction to prioritize the more constrained V view, quantizing each latent row as one block with a scale $a_t>0$:
\begin{equation}
 Z_{t,j}=\Qfour\!\left(X_{t,j}/a_t\right),
 \label{eq:pvquant}
\end{equation}
where $\Qfour$ is saturating nearest-even E2M1 rounding. With $Z$ fixed, each K scale is obtained by rounding the least-squares fit to the original $X$:
\begin{equation}
 k_{t,c}=\Pi_{\mathcal F_8^+}\!\left(
 \frac{\sum_{j\in\mathcal B_c}X_{t,j}Z_{t,j}}
 {\sum_{j\in\mathcal B_c}Z_{t,j}^2}\right).
 \label{eq:keyfit}
\end{equation}
Here $\mathcal B_c=\{16c,\ldots,16c+15\}$ and $\Pi_{\mathcal F_8^+}$ rounds to finite nonnegative E4M3. Zero rows and zero-energy blocks use $a_t=1$ and $k=0$, respectively. Fitting $k$ leaves $Z$ and the V view unchanged.

A single KV cache write kernel produces the FP4 payload $Z$ and its V scale $a$, fits the K scales $k$, and stores all three in the layouts required by the decode kernel. It encodes prompt tokens in parallel during prefill. During decoding, new tokens are encoded and appended without requantizing or modifying existing cache entries. For $d=512$ and $r=64$, the cache occupies $d/2+d/16+4+r=356$ B/token: 256 code bytes, 32 K-scale bytes, four V-scale bytes, and 64 RoPE bytes. It uses 38.2\% less storage than FP8 and only four more bytes per token than CAKE's NVFP4 KV cache.

\begin{figure}[t]
\centering
\includegraphics[width=\linewidth]{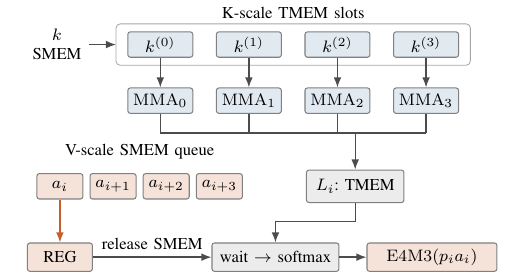}
\caption{Consumer-specific scale delivery. Four K-scale TMEM slots remove overwrite waits between QK MMAs. V scales reach registers before waiting for scores, freeing SMEM slots for prefetching while register copies feed probability conversion. Here $a_i$ is the token-scale vector for KV tile $i$. REG denotes registers.}
\label{fig:scale-pipeline}
\end{figure}

\subsection{Hardware-Efficient FP4 MLA Decoding}
\label{sec:kernel}
\label{sec:scale-delivery}
\textbf{Native QK and Mixed-Precision PV.}
For each cache tile, QK multiplies NVFP4 content queries $Q_c$ by the K view of $Z$ through native block-scaled MMA. A separate E4M3 product adds the RoPE contribution to FP32 scores before masking and softmax. V scales enter probability conversion instead. Online softmax~\cite{onlinesoftmax} updates the running maximum $m$, denominator $\ell$, and output accumulator $O$ as
\begin{align}
 m'_h&=\max(m_h,\max_tL_{h,t}),&
 \alpha_h&=e^{m_h-m'_h},\notag\\
 p_{h,t}&=e^{L_{h,t}-m'_h},&
 \ell'_h&=\alpha_h\ell_h+\sum_tp_{h,t},\label{eq:online}\\
 \widetilde p_{h,t}&=\Qeight(p_{h,t}a_t),&
 O'_h&=\alpha_hO_h+\widetilde p_hZ.\notag
\end{align}
Here $p$ denotes unnormalized weights within the tile, and $\Qeight$ rounds to E4M3. V scaling changes only the numerator, while the denominator sums $p$, not $pa$. To implement Eq.~\eqref{eq:valuefold}, the kernel scales $p_{h,t}$ by $a_t$ using packed FP32 multiplication, then converts the result to E4M3.

\begin{figure*}[t]
\centering
\raisebox{-61.48068pt}{\includegraphics[width=\linewidth]{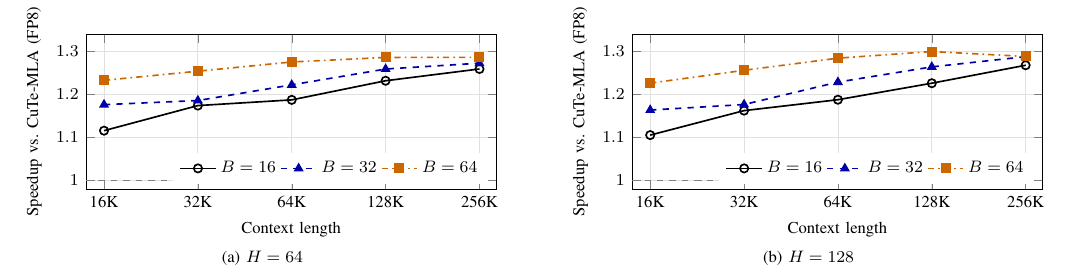}}
\caption{AscMLA speedup over FP8 CuTe-MLA under cold-L2 CUDA Graph replay.}
\label{fig:perf}
\end{figure*}

\begin{table*}[t]
\caption{Decode latency ($\mu$s) at representative workloads, $H=128$.}
\label{tab:latency}
\centering
\begin{tabular*}{0.9\textwidth}{@{\extracolsep{\fill}}lcccccr@{}}
\toprule
$B\times T$ & CuTe-MLA & TRT-MLA & Dequant+CuTe & Cake-MLA & Sage-MLA & AscMLA\\
Latent & FP8 & FP8 & NVFP4 & NVFP4 & $2\times$NVFP4 & Asym. FP4\\
\midrule
$16\times16$K & 47.30 & 53.18 & 77.59 & 46.91 & 47.82 & \textbf{42.78}\\
$32\times16$K & 76.98 & 85.58 & 137.22 & 75.23 & 79.98 & \textbf{66.14}\\
$32\times64$K & 237.38 & 259.36 & 478.23 & 238.18 & 236.87 & \textbf{193.17}\\
$64\times64$K & 456.10 & 492.02 & 940.89 & 449.28 & 425.63 & \textbf{355.06}\\
$64\times256$K & 1874.40 & 2057.70 & 4070.44 & 1793.86 & 1614.55 & \textbf{1454.64}\\
$128\times256$K & 4524.98 & 4868.11 & 8628.12 & 4016.54 & 3614.19 & \textbf{3514.46}\\
\bottomrule
\end{tabular*}
\end{table*}

The PV product uses MXFP8--MXFP4 MMA with E4M3 $\widetilde p$ and E2M1 $Z$, avoiding online NVFP4 probability quantization. The MX form can also apply fixed V-channel scaling directly in PV~\cite{mxformats}. Its scale operands occupy four reserved TMEM columns and are staged once for reuse. After the final tile, normalization returns BF16 $Y_h=O_h/\ell_h$.

\textbf{Consumer-Specific Scale Delivery.}
The two scale views require different buffering schedules. K scales remain live until their MMA completes, whereas V scales are consumed by compute warps during probability conversion. Fig.~\ref{fig:scale-pipeline} illustrates how AscMLA uses these distinct lifetimes to overlap scale delivery with computation.

The warp-specialized implementation draws on FlashInfer's FP8 CuTe DSL MLA design~\cite{flashinfer}. Each cooperative thread array (CTA) in a two-CTA cluster uses twelve warps: four compute, four correction, one MMA, two payload loaders, and one scale producer. Tensor Memory Accelerator (TMA) loads the shared global codes into distinct K/V SMEM layouts.

QK processes each tile in four 128-channel segments. Reusing one TMEM scale slot forces each overwrite to wait for the preceding MMA, even when the next payload is ready. K-scale multibuffering removes this dependency with four disjoint slots within the existing TMEM allocation. TMA transfers prepacked Q/K scales to SMEM. The MMA warp copies K-scale slices to their slots before issuing segment MMAs. Q scales persist across tiles.

V-scale prefetching uses an independent four-stage SMEM queue for FP32 token scales. Each producer lane issues a 128-bit \texttt{cp.async} for four adjacent scales. Compute warps load scales from a ready stage into registers and release its SMEM slot before waiting for the corresponding scores. The producer can then refill the slot while register copies serve probability conversion across heads.

Compute warps publish E4M3 probabilities to SMEM, the MMA warp accumulates PV in TMEM, and correction warps rescale and normalize outputs. Split-KV supplies parallelism at small batches~\cite{flashinfer}, with dependent launch coordinating partial-output reduction.

\section{Evaluation}
\label{sec:evaluation}
We evaluate AscMLA's performance, memory savings, and accuracy tradeoffs. Kernel experiments compare FP8 and FP4 baselines. Ablations isolate the effects of reconstruction order and scale delivery. End-to-end experiments assess serving throughput under memory and latency constraints.

\subsection{Experimental Setup}
We use an eight-GPU platform with CUDA compute capability 10.3 and CUDA 13.0 for our experiments.
The software stack includes CUTLASS DSL 4.6.2, Triton 3.7.1, PyTorch 2.13.0, FlashInfer 0.6.18, and SGLang 0.5.21~\cite{sglang}.

We compare AscMLA with two FP8 methods and three FP4 methods. The FP8 methods are FlashInfer's CuTe DSL FP8 MLA implementation~\cite{flashinfer} (CuTe-MLA) and TRT-LLM's FP8 MLA kernel~\cite{trtllm} (TRT-MLA). The FP4 methods include CuTe-MLA with external NVFP4-to-FP8 dequantization (Dequant+CuTe), CAKE's NVFP4 MLA kernel~\cite{cake} (Cake-MLA), and an MLA adaptation of SageAttention3's separate K/V encoding~\cite{sageattention3} (Sage-MLA). We compile Cake-MLA separately with CUDA 13.4 to enable QMUL4\footnote{\url{https://github.com/flashinfer-ai/flashinfer/pull/6242}.}. All methods share source tensors, and all FP4 payloads use amax-based quantization at their respective scale granularities. AscMLA uses an asymmetrically scaled FP4 format (Asym. FP4).

\subsection{Decode Performance}
We benchmark forty workload configurations on a single GPU, with batch size $B\in\{16,32,64,128\}$, context length $T\in\{16,32,64,128,256\}$K, and query-head count $H\in\{64,128\}$. All kernel-level experiments use query length one, 512-dimensional latents, 64-dimensional E4M3 RoPE, 64-token pages, and BF16 output. Performance inputs are generated from Gaussian FP8 tensors.

We time complete decode execution with CUPTI during cold-L2 CUDA Graph replay. Each replay starts after an untimed L2 flush. Queries and KV caches are prepared in advance. The timed region includes attention, split reduction, and per-call external dequantization where applicable. We report the median of five per-round medians.

Table~\ref{tab:latency} reports latencies for six representative workloads at $H=128$. Across all forty configurations, AscMLA achieves geometric-mean speedups of 1.24$\times$ over CuTe-MLA, 1.35$\times$ over TRT-MLA, 2.49$\times$ over Dequant+CuTe, 1.26$\times$ over Cake-MLA, and 1.21$\times$ over Sage-MLA.

Fig.~\ref{fig:perf} shows the speedup over CuTe-MLA by context length and batch size. At $B=16$, the speedup grows from about 1.11$\times$ at 16K to 1.26--1.27$\times$ at 256K. At larger batches, the speedup varies less with context length, staying within 1.23--1.30$\times$ at $B=64$ and showing a similar plateau at $B=128$.

Among the FP4 baselines, Sage-MLA's native NVFP4 PV MMA offers higher peak throughput~\cite{nvfp4}, but online NVFP4 quantization of $P$ offsets the benefit. Cake-MLA fuses V dequantization into attention and keeps the expanded V in SMEM, bringing decode latency close to that of the FP8 baselines. Dequant+CuTe instead expands the cache to FP8 in a separate kernel and writes it to global memory, resulting in the highest decode latency. AscMLA operates directly on a shared FP4 payload and achieves the lowest decode latency at every point in the benchmark grid.

\subsection{Attention Accuracy}
We evaluate attention accuracy on five FP8 latent distributions: RMS-normalized Gaussian inputs, ordered and shuffled channel gains (0.5--2), sparse outliers (0.2\% of entries amplified eightfold before normalization), and correlated inputs that share a Gaussian component within 16-token groups. We combine three seeds with $T\in\{1,4,16\}$K and $H\in\{64,128\}$, yielding 90 equally weighted cases.

We measure attention-output error as $\|\widehat A-A\|_F/\|A\|_F$, where $A$ is the output of FP8 CuTe-MLA. The comparison includes operand conversions. Native NVFP4 QK paths quantize content queries to NVFP4, while the other paths use E4M3. K/V latent errors use the original FP8 latent as the reference. Accuracy runs use one KV split except for TRT-MLA, which uses its own dispatcher.

Table~\ref{tab:accuracy} reports mean attention-output error and persistent KV cache storage. Both FP8 baselines use 576 B/token, compared with 352--356 B/token for the single-payload FP4 methods and 640 B/token for Sage-MLA. Dequant+CuTe performs QK in FP8 after expansion and has the lowest error among the FP4 methods. Its temporary FP8 buffer adds 576 B/token during decoding, bringing the combined cache and buffer storage to 928 B/token. Cake-MLA achieves 11.88\% error with block-scaled V and FP8 PV. AscMLA uses 44.4\% less KV cache storage than Sage-MLA, with a mean error of 13.37\% versus 12.66\%. Their mean cosine similarities are 0.9908 and 0.9920, respectively.

\begin{figure}[t]
\centering
\includegraphics[width=\linewidth]{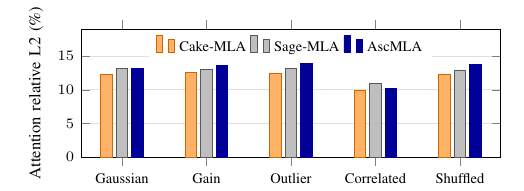}
\caption{Mean attention-output error across five input distributions.}
\label{fig:accuracy}
\end{figure}

\begin{table}[t]
\caption{Persistent KV cache storage per token (including RoPE) and mean attention-output error relative to FP8 CuTe-MLA \\ over 90 configurations.}
\label{tab:accuracy}
\centering\small
\setlength{\tabcolsep}{4pt}
\begin{tabular}{@{}lrr@{}}
\toprule
Method & KV cache (B/token) & Attention L2 (\%) $\downarrow$\\
\midrule
CuTe-MLA & 576 & 0\\
TRT-MLA & 576 & 1.55\\
Dequant+CuTe & 352 & 6.69\\
Cake-MLA & 352 & 11.88\\
Sage-MLA & 640 & 12.66\\
\textbf{AscMLA} & 356 & 13.37\\
\bottomrule
\end{tabular}
\end{table}

Fig.~\ref{fig:accuracy} shows how attention-output error varies across input distributions. Compared with Sage-MLA, AscMLA has similar error on Gaussian inputs, lower error on correlated inputs, and higher error on sparse-outlier inputs. A few large entries can set the amax scale for an entire token, making per-token quantization sensitive to outliers. Cake-MLA has lower error than both methods in each distribution.

AscMLA offers a favorable accuracy–performance tradeoff. Compared with Sage-MLA, it achieves a 1.21$\times$ geometric-mean decode speedup while increasing mean attention-output error by 0.71 percentage points.

\subsection{Ablation Studies}
\label{sec:order-ablation}
\textbf{Reconstruction order.}
We evaluate two ablations of AscMLA's cache construction on the same 90 cases without changing the decode kernel. AscMLA without K-scale fitting keeps the PV-first payload and repeats each rounded V scale across the corresponding K blocks. AscMLA with QK-first construction chooses the shared codes by amax-based NVFP4 quantization of K, then fits one V scale per token. We measure relative L2 error for the K/V latents and attention output.

Relative to full AscMLA, omitting K-scale fitting raises mean K latent and attention-output errors by 7.5\% and 6.1\%, respectively, without changing V latent error. QK-first construction lowers K latent error by 24.4\% but raises V latent and attention-output errors by 138.0\% and 87.9\%, respectively.

\textbf{Scale delivery.}
We independently ablate the two scale-delivery mechanisms in Section~\ref{sec:scale-delivery}. The K-scale variant reuses a single TMEM slot, while the V-scale variant reduces the SMEM queue from four stages to one. Both keep arithmetic, TMEM allocation, warp/register budgets, and split policy fixed and produce bit-identical outputs. Table~\ref{tab:pipeline} reports decode speedup relative to full AscMLA.

\begin{table}[t]
\caption{Decode speedup for scale-delivery ablations at $H=128$ \\ (full AscMLA $=1\times$).}
\label{tab:pipeline}
\centering\small
\begin{tabular}{@{}lrr@{}}
\toprule
$B\times T$ & \shortstack{w/o K-scale\\multibuffering} & \shortstack{w/o V-scale\\multistage prefetching}\\
\midrule
$16\times16$K & 0.90$\times$ & 0.89$\times$\\
$32\times64$K & 0.77$\times$ & 0.87$\times$\\
$64\times256$K & 0.69$\times$ & 0.88$\times$\\
\bottomrule
\end{tabular}
\end{table}

Omitting K-scale multibuffering has the largest effect at $64\times256$K, where speedup drops to 0.69$\times$. The single slot cannot be overwritten until the preceding QK MMA completes, even when the next payload is ready. Without multistage V-scale prefetching, speedups range from 0.87$\times$ to 0.89$\times$ because the shorter queue limits how far scale loading can advance ahead of score consumption.

At $64\times256$K, Nsight Compute shows lower tensor-pipeline activity in the main attention kernel for both ablations, with unchanged tensor-active cycle counts. The drop is consistent with reduced overlap between scale delivery and computation.

\subsection{End-to-End Evaluation}
\textbf{Serving throughput.}
We implement AscMLA as an extension to FlashInfer 0.6.18. We then integrate the kernel as a decode backend in SGLang 0.5.21 and adapt KV cache allocation, writes, and transfers to support the asymmetric FP4 format. We deploy ModelOpt-NVFP4 Kimi-K2.6 on four prefill and four decode GPUs, using Mooncake~\cite{mooncake} for KV transfer. Both sides are configured with TP=4. Decode enables four-way data-parallel attention and CUDA Graph. Prefill uses TRT-MLA, while the decode backend varies across all six methods without speculative decoding. Prefill/decode memory fractions are 0.7/0.9, and decode concurrency is capped at 288.

We measure serving throughput for each decode backend and determine whether it meets both service-level objectives (SLOs). The SLOs require a P90 time to first token (TTFT) of at most 20\,s and a mean time per output token (TPOT) of at most 50\,ms. We test arrival rates of 0.9, 1.2, 1.5, and 1.8 requests/s. At each rate, all methods use the same 120\,s open-loop arrival schedule. Each request uses a 64K input derived from ShareGPT~\cite{sharegpt} and generates 4K output tokens. We warm the prefill-side cache with the test prompts and disable decode-side radix caching. We report output tokens per second over the full run, including drain. Latency statistics exclude arrivals in the first 30\,s.

\begin{figure}[t]
\centering
\includegraphics[width=\linewidth]{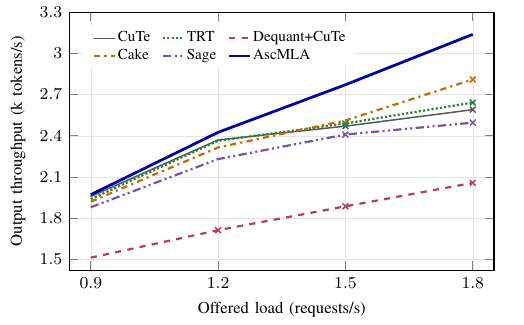}
\caption{Kimi-K2.6 serving throughput with 64K input and 4K output tokens per request. Crosses mark SLO violations.}
\label{fig:e2e}
\end{figure}

Fig.~\ref{fig:e2e} shows serving throughput and the request rates at which SLO violations first occur. Table~\ref{tab:serving-slo} reports TTFT and TPOT at the highest load. The two FP8 backends and Sage-MLA fill their decode KV pools at roughly 140 active requests, forcing new requests to wait on the prefill side for decode slots. Limited decode concurrency keeps mean TPOT relatively low, but all three methods first miss the P90 TTFT target at 1.5 requests/s. The two FP8 backends deliver similar throughput, but Sage-MLA trails both because its cache writes require additional V quantization. Dequant+CuTe avoids the admission queue with its NVFP4 cache, but its separate dequantization kernel slows decode, and mean TPOT first exceeds its target at 1.2 requests/s.

\begin{table}[t]
\caption{Serving latency at 1.8 requests/s (SLO violations in bold).}
\label{tab:serving-slo}
\centering\small
\setlength{\tabcolsep}{5pt}
\begin{tabular}{@{}lrr@{}}
\toprule
Method & P90 TTFT (s) & Mean TPOT (ms)\\
\midrule
CuTe-MLA & \textbf{89.79} & 42.25\\
TRT-MLA & \textbf{82.79} & 40.82\\
Dequant+CuTe & 1.00 & \textbf{80.98}\\
Cake-MLA & 0.77 & \textbf{52.66}\\
Sage-MLA & \textbf{98.40} & 40.53\\
\textbf{AscMLA} & 1.11 & 44.05\\
\bottomrule
\end{tabular}
\end{table}

Cake-MLA uses the same NVFP4 cache but fuses V dequantization into the attention kernel. At low load, the extra cost of NVFP4 cache writes outweighs its decode savings, leaving Cake-MLA slightly behind FP8 in throughput. Once the FP8 pools fill, Cake-MLA leads the baselines in throughput and is the only one to meet both SLOs at 1.5 requests/s. Its mean TPOT first exceeds the limit at 1.8 requests/s. AscMLA consumes V directly in FP4 and meets both SLOs throughout the sweep. At the highest load, it delivers 3139 output tokens/s, 18.8\% above TRT-MLA and 11.7\% above Cake-MLA.

\textbf{Downstream accuracy.}
We evaluate Kimi-K2.6~\cite{kimik26} with ModelOpt-NVFP4 weights and Kimi-K3~\cite{kimik3} with MXFP4 weights, using AscMLA for MLA decoding. The two models score 89.4\% and 92.8\% on GPQA-Diamond~\cite{gpqa}, averaged over eight rounds each, compared with officially reported accuracies of 90.5\% and 93.5\%.

\section{Related Work}
\textbf{KV cache compression.}
MLA compresses K and V into a shared latent~\cite{deepseekv2,deepseekv3}. ClusterKV selects token clusters for attention~\cite{clusterkv}, MILLION encodes cache entries with product quantization~\cite{million}, and KIVI quantizes K per channel and V per token~\cite{kivi}. AscMLA stores one block-scaled FP4 payload per MLA latent, with channel-block K and per-token V scales.

\textbf{Low-precision attention.}
FlashAttention-3/4 overlap data movement and computation in asynchronous pipelines for FP8 attention~\cite{fa3,fa4}. BitDecoding tailors low-bit KV layouts to Tensor Cores~\cite{bitdecoding}, while MpFA pairs NVFP4 QK with FP8 PV~\cite{mpfa}. BitDecoding-FP4's MLA path dequantizes V to FP16~\cite{bitdecodingfp4}, while CAKE's NVFP4 kernel converts it to FP8~\cite{cake}. AscMLA folds V scales into probabilities as in SnapMLA~\cite{snapmla}, so PV uses the shared FP4 codes directly.

\section{Conclusion}
AscMLA is a representation and kernel co-design for MLA decoding with asymmetrically scaled FP4 KV caches. It uses 38.2\% less cache memory than FP8 and achieves geometric-mean speedups of 1.35$\times$ over TRT-LLM's FP8 and 1.26$\times$ over CAKE's NVFP4 MLA kernels. When used as the decode backend for Kimi-K2.6 serving at 1.8 requests/s, AscMLA meets both latency targets and delivers 18.8\% higher throughput than the TRT-LLM FP8 baseline.

\nobalance
\bibliographystyle{IEEEtran}
\bibliography{references}

@inproceedings{kivi,
  author    = {Liu, Zirui and Yuan, Jiayi and Jin, Hongye and Zhong, Shaochen and Xu, Zhaozhuo and Braverman, Vladimir and Chen, Beidi and Hu, Xia},
  title     = {{KIVI: A tuning-free asymmetric 2bit quantization for KV cache}},
  booktitle = {Proceedings of the 41st International Conference on Machine Learning (ICML)},
  volume    = {235},
  pages     = {32332--32344},
  year      = {2024},
}

@inproceedings{fa3,
  author    = {Shah, Jay and Bikshandi, Ganesh and Zhang, Ying and Thakkar, Vijay and Ramani, Pradeep and Dao, Tri},
  title     = {{FlashAttention-3: Fast and accurate attention with asynchrony and low-precision}},
  booktitle = {Advances in Neural Information Processing Systems (NeurIPS)},
  volume    = {37},
  pages     = {68658--68685},
  year      = {2024},
  doi       = {10.52202/079017-2193},
  note      = {{doi}: 10.52202/079017-2193}
}

@inproceedings{flashinfer,
  author    = {Ye, Zihao and Chen, Lequn and Lai, Ruihang and Lin, Wuwei and Zhang, Yineng and Wang, Stephanie and Chen, Tianqi and Kasikci, Baris and Grover, Vinod and Krishnamurthy, Arvind and Ceze, Luis},
  title     = {{FlashInfer: Efficient and customizable attention engine for LLM inference serving}},
  booktitle = {Proceedings of Machine Learning and Systems (MLSys)},
  volume    = {7},
  year      = {2025},
}

@misc{nvfp4,
  author       = {{NVIDIA}},
  title        = {{Pretraining large language models with NVFP4}},
  howpublished = {arXiv preprint arXiv:2509.25149},
  year         = {2026},
}

@misc{mpfa,
  author       = {Deng, Chencheng and Fang, Jianbin and Dong, Dezun},
  title        = {{MpFA: Hardware-efficient train-free QK4V8 FlashAttention kernels on Blackwell GPUs}},
  howpublished = {arXiv preprint arXiv:2609.33135},
  year         = {2026},
}

@misc{deepseekv2,
  author       = {{DeepSeek-AI}},
  title        = {{DeepSeek-V2: A strong, economical, and efficient mixture-of-experts language model}},
  howpublished = {arXiv preprint arXiv:2405.04434},
  year         = {2024},
}

@misc{deepseekv3,
  author       = {{DeepSeek{-}AI}},
  title        = {{DeepSeek-V3 technical report}},
  howpublished = {arXiv preprint arXiv:2412.19437},
  year         = {2024},
}

@article{roformer,
  author  = {Su, Jianlin and Ahmed, Murtadha and Lu, Yu and Pan, Shengfeng and Bo, Wen and Liu, Yunfeng},
  title   = {{RoFormer: Enhanced transformer with Rotary Position Embedding}},
  journal = {Neurocomputing},
  volume  = {568},
  pages   = {127063},
  year    = {2024},
  doi     = {10.1016/j.neucom.2023.127063},
  note    = {{doi}: 10.1016/j.neucom.2023.127063}
}

@misc{trtllm,
  author       = {{NVIDIA}},
  title        = {{TensorRT-LLM}},
  url          = {https://github.com/NVIDIA/TensorRT-LLM}
}

@misc{cake,
  author       = {Ye, Zihao and Huang, Yingyi and Jin, Hongyi and Hou, Bohan and Shao, Junru and Yu, Zhongming and Chen, Jinqi and Cowan, Meghan and Cao, Shiyi and Xing, Shanli and Chen, Hanfeng and Grover, Vinod and Chen, Tianqi and Ceze, Luis},
  title        = {{CAKE: Compiler-agent co-design for frontier kernel evolution}},
  howpublished = {arXiv preprint arXiv:2608.12629},
  year         = {2026},
}

@inproceedings{transformer,
  author    = {Vaswani, Ashish and Shazeer, Noam and Parmar, Niki and Uszkoreit, Jakob and Jones, Llion and Gomez, Aidan N. and Kaiser, {\L}ukasz and Polosukhin, Illia},
  title     = {{Attention is all you need}},
  booktitle = {Advances in Neural Information Processing Systems (NeurIPS)},
  volume    = {30},
  pages     = {5998--6008},
  year      = {2017},
}

@inproceedings{sageattention3,
  author    = {Zhang, Jintao and Wei, Jia and Wang, Haoxu and Zhang, Pengle and Xu, Xiaoming and Huang, Haofeng and Jiang, Kai and Chen, Jianfei and Zhu, Jun},
  title     = {{SageAttention3: Microscaling FP4 attention for inference and an exploration of 8-bit training}},
  booktitle = {Advances in Neural Information Processing Systems (NeurIPS)},
  volume    = {38},
  pages     = {53901--53931},
  year      = {2025},
  doi       = {10.52202/085713-1799},
  note      = {{doi}: 10.52202/085713-1799}
}

@misc{bitdecodingfp4,
  author       = {Du, Dayou and Ng, Woon Yee and Sheng, Guangming and Lau, Nick and Cheng, Jianyi and Mai, Luo},
  title        = {{BitDecoding-FP4: Native FP4 KV cache decoding via algorithm-system co-design}},
  howpublished = {arXiv preprint arXiv:2611.09825},
  year         = {2026},
}

@inproceedings{fa4,
  author    = {Zadouri, Ted and Hoehnerbach, Markus and Shah, Jay and Thakkar, Vijay and Dao, Tri},
  title     = {{FlashAttention-4: Algorithm and kernel pipelining co-design for asymmetric hardware scaling}},
  booktitle = {Proceedings of Machine Learning and Systems (MLSys)},
  volume    = {8},
  pages     = {912--926},
  year      = {2026},
}

@misc{snapmla,
  author       = {Zhang, Yifan and Su, Zunhai and Hu, Shuhao and Yang, Rui and Wu, Wei and Qian, Yulei and Xie, Yuchen and Cai, Xunliang},
  title        = {{SnapMLA: Efficient long-context MLA decoding via hardware-aware FP8 quantized pipelining}},
  howpublished = {arXiv preprint arXiv:2602.10718},
  year         = {2026},
}

@misc{onlinesoftmax,
  author       = {Milakov, Maxim and Gimelshein, Natalia},
  title        = {{Online normalizer calculation for softmax}},
  howpublished = {arXiv preprint arXiv:1805.02867},
  year         = {2018},
}

@misc{mxformats,
  author       = {Darvish Rouhani, Bita and Zhao, Ritchie and More, Ankit and Hall, Mathew and Khodamoradi, Alireza and Deng, Summer and Choudhary, Dhruv and Cornea, Marius and Dellinger, Eric and Denolf, Kristof and Stosic, Dusan and Elango, Venmugil and Golub, Maximilian and Heinecke, Alexander and James-Roxby, Phil and Jani, Dharmesh and Kolhe, Gaurav and Langhammer, Martin and Li, Ada and Melnick, Levi and Mesmakhosroshahi, Maral and Rodriguez, Andres and Schulte, Michael and Shafipour, Rasoul and Shao, Lei and Siu, Michael and Dubey, Pradeep and Micikevicius, Paulius and Naumov, Maxim and Verrilli, Colin and Wittig, Ralph and Burger, Doug and Chung, Eric},
  title        = {{Microscaling data formats for deep learning}},
  howpublished = {arXiv preprint arXiv:2310.10537},
  year         = {2023},
}

@inproceedings{vllm,
  author    = {Kwon, Woosuk and Li, Zhuohan and Zhuang, Siyuan and Sheng, Ying and Zheng, Lianmin and Yu, Cody Hao and Gonzalez, Joseph E. and Zhang, Hao and Stoica, Ion},
  title     = {{Efficient memory management for large language model serving with PagedAttention}},
  booktitle = {Proceedings of the 29th Symposium on Operating Systems Principles (SOSP)},
  pages     = {611--626},
  year      = {2023},
  doi       = {10.1145/3600006.3613165},
  note      = {{doi}: 10.1145/3600006.3613165}
}

@inproceedings{sglang,
  author    = {Zheng, Lianmin and Yin, Liangsheng and Xie, Zhiqiang and Sun, Chuyue and Huang, Jeff and Yu, Cody Hao and Cao, Shiyi and Kozyrakis, Christos and Stoica, Ion and Gonzalez, Joseph E. and Barrett, Clark and Sheng, Ying},
  title     = {{SGLang: Efficient execution of structured language model programs}},
  booktitle = {Advances in Neural Information Processing Systems (NeurIPS)},
  volume    = {37},
  pages     = {62557--62583},
  year      = {2024},
  doi       = {10.52202/079017-2000},
  note      = {{doi}: 10.52202/079017-2000}
}

@inproceedings{mooncake,
  author    = {Qin, Ruoyu and Li, Zheming and He, Weiran and Cui, Jialei and Ren, Feng and Zhang, Mingxing and Wu, Yongwei and Zheng, Weimin and Xu, Xinran},
  title     = {{Mooncake: Trading more storage for less computation---A KVCache-centric architecture for serving LLM chatbot}},
  booktitle = {Proceedings of the 23rd USENIX Conference on File and Storage Technologies (FAST)},
  pages     = {155--170},
  year      = {2025},
}

@misc{kimik26,
  author       = {{Kimi~Team}},
  title        = {{Kimi K2.6: Advancing open-source coding}},
  year         = {2026},
  url          = {https://www.kimi.com/en/blog/kimi-k2-6}
}

@misc{kimik3,
  author       = {{Kimi Team}},
  title        = {{Kimi K3: Open frontier intelligence}},
  howpublished = {arXiv preprint arXiv:2607.24653},
  year         = {2026},
}

@inproceedings{gpqa,
  author    = {Rein, David and Hou, Betty Li and Cooper Stickland, Asa and Petty, Jackson and Pang, Richard Yuanzhe and Dirani, Julien and Michael, Julian and Bowman, Samuel R.},
  title     = {{GPQA: A graduate-level Google-proof Q\&A benchmark}},
  booktitle = {Proceedings of the First Conference on Language Modeling (COLM)},
  year      = {2024},
}

@inproceedings{clusterkv,
  author    = {Liu, Guangda and Li, Chengwei and Zhao, Jieru and Zhang, Chenqi and Guo, Minyi},
  title     = {{ClusterKV: Manipulating LLM KV cache in semantic space for recallable compression}},
  booktitle = {Proceedings of the 62nd ACM/IEEE Design Automation Conference (DAC)},
  pages     = {1--7},
  year      = {2025},
  doi       = {10.1109/DAC63849.2025.11132479},
  note      = {{doi}: 10.1109/DAC63849.2025.11132479}
}

@inproceedings{million,
  author    = {Wang, Zongwu and Xu, Peng and Liu, Fangxin and Hu, Yiwei and Sun, Qingxiao and Li, Gezi and Li, Cheng and Wang, Xuan and Jiang, Li and Guan, Haibing},
  title     = {{MILLION: Mastering long-context LLM inference via outlier-immunized KV product quantization}},
  booktitle = {Proceedings of the 62nd ACM/IEEE Design Automation Conference (DAC)},
  pages     = {1--7},
  year      = {2025},
  doi       = {10.1109/DAC63849.2025.11132862},
  note      = {{doi}: 10.1109/DAC63849.2025.11132862}
}

@inproceedings{bitdecoding,
  author    = {Du, Dayou and Cao, Shijie and Cheng, Jianyi and Mai, Luo and Cao, Ting and Yang, Mao},
  title     = {{BitDecoding: Unlocking Tensor Cores for long-context LLMs with low-bit KV cache}},
  booktitle = {Proceedings of the IEEE International Symposium on High Performance Computer Architecture (HPCA)},
  pages     = {1--13},
  year      = {2026},
  doi       = {10.1109/HPCA68181.2026.11408481},
  note      = {{doi}: 10.1109/HPCA68181.2026.11408481}
}

@misc{sharegpt,
  author       = {{ShareGPT}},
  title        = {{ShareGPT}},
  url          = {https://github.com/domeccleston/sharegpt}
}
\end{document}